\documentclass[3p,preprint]{elsarticle}

\usepackage{graphicx}
\usepackage{amssymb}
\usepackage{hyperref}
\usepackage{braket}
\usepackage{amsmath}
\usepackage{xcolor}
\usepackage{soul}
\usepackage{bm}
\setstcolor{red}
\setul{}{1.5pt}
\usepackage[normalem]{ulem}
\graphicspath{{./figs/}}

\newcommand{\espin} {\texttt{ESpinS}}
\newcommand{\beq} {\begin{equation}}
\newcommand{\eeq} {\end{equation}}
\newcommand{\dm}  {Dzyaloshinskii-Moriya}
\newcommand\redsout{\bgroup\markoverwith{\textcolor{red}{\rule[0.5ex]{2pt}{1.4pt}}}\ULon}
\newcounter{bla}

\journal{Computational Materials Science}

\begin{document}

\begin{frontmatter}

\title{\espin: A program for classical Monte-Carlo simulations of spin systems}
\author[a]{Nafise Rezaei}
\author[a]{Mojtaba Alaei\corref{author}}
\author[a]{Hadi Akbarzadeh}

\cortext[author] {Corresponding author.\\\textit{E-mail address:}  m.alaei@iut.ac.ir}
\address[a]{Department of Physics, Isfahan University of Technology, Isfahan 84156-83111, Iran}

\begin{abstract}
We present \espin\ (Esfahan Spin Simulation) package to evaluate the thermodynamic properties 
of spin systems described by a spin model Hamiltonian.
In addition to the Heisenberg exchange term, the spin Hamiltonian can contain
interactions such as bi-quadratic,
 Dzyaloshinskii-Moriya, and  single-ion anisotropy.
By applying the classical Monte-Carlo simulation, \espin\
simulates the behavior of spin systems versus temperature. 
\espin\ ables to calculate the specific heat, susceptibility, 
staggered magnetization, energy histogram,
fourth-order Binder cumulants, and the neutron scattering structure factor.
Further, it can compute the user-defined magnetic 
order parameter {\it i.e.} summation of projection of spins on the user-defined directions and the 
physical quantities based on it.
\espin\ works by either local update algorithm or parallel tempering algorithm. 
The latter feature is an appropriate option
 for considering the frustrated and spin glass magnetic systems.
\espin\ is written in Fortran 90 and can be run in single or parallel mode. 
The package is freely available under
the GPL license  (see {\tt https://github.com/nafiserb/ESpinS}).

\end{abstract}


\end{frontmatter}


\section{Introduction}
\label{}
Nowadays, with the extensive use of magnetic materials, the critical phenomena of spin systems 
have attracted much attention in both theoretical and experimental fields. 
One of the challenges in the theoretical study of phase transition is the massive number of particles in the systems. 
On the other hand, the exact analytical solutions can be applied to only a few simple spin models in one or two-dimensional systems~\cite{plischke1994}. 
Hence the computational approaches are required to make a bridge between the theoretical models and reality.

Monte-Carlo (MC) simulations are computational techniques, 
which assist in a better understanding of systems with a large number of interacting particles, 
where problems involve integrations over broad phase space\cite{MC1,MC2}. 
So with ongoing advances in computer technology, 
the MC simulations help gain more accurate physical insight into the real materials. 
The basic idea of MC simulation for solving an integration problem is simple: 
the random sampling of phase space instead of the regular sampling, 
at the expense of statistical error. 
The superiority of the MC simulations over ordinary numerical methods is 
that statistical error is independent of the phase space dimensions. 
The only difficulty of MC simulations is how to select the stochastic sampling 
to reduce the statistical errors. 
A procedure widely used for stochastic sampling is the local update Metropolis algorithm. 
Thence, the MC simulation community offered several techniques for the improvement of local update Metropolis schemes. 
Some approaches are general methods like parallel tempering algorithm~\cite{ptmc1,ptmc2,ptmc3}, 
while some work for specific cases, like Swendsen-Wang~\cite{swendsen-wang} and Wolff cluster update algorithms~\cite{wolff}.

To gain macroscopic insight into the properties of the magnetic material, 
such as phase transition and stable magnetic configuration,  
a spin model Hamiltonian consists of the microscopic magnetic interactions 
like Heisenberg exchange parameters are needed.
 The microscopic interactions can be derived from experiments like the 
inelastic neutron scattering~\cite{Pepy1974,INS_NiO,SNWO,plumb2019} or first-principles calculations 
like density functional theory (DFT)~\cite{DFT_MnO,Heusler,Marjana,Kresse}.
 With the investigation of spin models through a 
classical spin MC simulation package, one can estimate the expected value of thermodynamic quantities.
Unlike the high levels of availability for the first-principles packages, 
there are very few free available classical MC simulation packages such as UppASD~\cite{uppasd}, VAMPIRE~\cite{vampire}, and SpinW~\cite{spinw}.
These MC simulation packages can fill the gap between microscopic and macroscopic results in magnetic material researches. 
Nevertheless, there is still room to develop MC simulation packages with distinct features. 
One of the essential elements for a MC package is to benefit from parallel computing 
to implement MC methods such as parallel tempering to reduce the MC simulation time.

In this work, we present our program, \espin, as an open-source classical spin MC software package. 
\espin\ enables calculating the thermodynamic properties 
and phase transition of magnetic materials based on both local updating and parallel tempering. 
With \espin, we can define an almost general spin model Hamiltonian that contains Heisenberg exchange, 
bi-quadratic, Dzyaloshinskii-Moriya, and single-ion anisotropy interactions.
Although the Heisenberg exchanges play a significant role in the magnetic thermodynamic properties, 
other interactions such as Dzyaloshinskii-Moriya and bi-quadratic interactions in frustrated systems may lead the
system to the single ground state~\cite{snwo_rezaei,Elhajal2005}. 
Recently, among these interactions, Dzyaloshinskii-Moriya gains more attraction due to skyrmions~\cite{skyrmions1,skyrmions2}. 
Using MC simulations, \espin\ gives the expectation values for
many desirable magnetic quantities such as susceptibility and static neutron scattering structure. 
We have tried to equip \espin\ with many features and enough flexibility. 
For example, the user can define order parameters, units, and even the local spin updating strategy. 
In the following, we explain more theoretical and technical details of \espin.

We organize the paper as follows. In Section \ref{theo}, we briefly review the 
MC simulations and spin model Hamiltonian used in \espin. 
After a brief discussion on the theoretical background of \espin, 
we describe its features, the details of its installation, and parallelization in Section \ref{program}. 
In Section \ref{exam}, we provide some examples and discuss the results.
\section{Theoretical background}
\label{theo}
\subsection{Spin model Hamiltonian}
In 1928, Heisenberg, taking into account Coulomb's repulsion and the Pauli exclusion principle, showed
that an effective Hamiltonian defined by $-\frac{1}{2}\sum_{i,j}J_{ij}{\mathbf{S}_i\cdot\mathbf{S}_j}$, can describe 
spin interactions in ferromagnets such as Fe, Co, and Ni~\cite{Heisenberg1928}.
Using the same principles, Anderson in 1951 indicated that this effective Hamiltonian is applicable 
for  systems such as MnO where magnetic ions like Mn$^{+2}$ interact with each other through 
intermediary non-magnetic ions such as Oxygens~\cite{Anderson1950}.  
This term is called  Heisenberg Hamiltonian, and interactions ($J_{ij}$) between magnetic moments ($\mathbf{S}_i$) 
are so-called exchange constants or parameters.
Although the Heisenberg Hamiltonian has a leading role in describing magnetic systems,
theoretically, the Hamiltonian can be extended to the following  spin model Hamiltonian~\cite{yosida1996,nolting2009}:
\beq
\label{ham}
\mathcal{H}=-\frac{1}{2}\sum_{i,j}J_{ij}{\mathbf{S}_i\cdot\mathbf{S}_j}+\frac{1}{2}\sum_{i,j}B_{ij}({\mathbf{S}_i\cdot\mathbf{S}_j})^2+
\frac{1}{2}\sum_{i,j}\mathbf{D}_{ij}\cdot(\mathbf{S}_i\times\mathbf{S}_j)
+\Delta\sum_i{(\hat{z_i}\cdot \mathbf{S}_i)^2},
\eeq
where $i$ and $j$ denote the sites of spins in the lattice. 
The second and third terms in Eq.~\ref{ham}
are so-call bi-quadratic and Dzyaloshinskii-Moriya (DM) interactions, respectively. 
The last term indicates the single-ion anisotropy interaction. 
$B_{ij}$ and  $\mathbf{D}_{ij}$ represent the strength of bi-quadratic and Dzyaloshinskii-Moriya (DM) 
interactions and $\Delta$ specifies (local) magnetic anisotropy energy.
Both single-ion and  DM terms are considered as anisotropic magnetic interactions. 
The former is due to either spin-orbit coupling to the crystalline electric field or dipole-dipole interactions~\cite{Schron2012}. 
The latter originates from the exchange interaction between the excited state of one ion, created by spin-orbit interaction, 
and the ground state of the other ion~\cite{yosida1996}. 
We can determine the direction of DM vectors($\mathbf{D}_{ij}$) by the Moriya rules~\cite{Moriya}.  
The local anisotropy direction of each site, shown by $\hat{z_i}$,
can be specified by first-principles calculations.
All of these terms are possible to be derived from quantum principles~\cite{yosida1996,nolting2009}. 
We should mention that it is still possible to extend the Hamiltonian by adding interactions such as four-spin exchange to the Hamiltonian.  
However, at the moment, \espin\ includes the terms presented in  Eq.~\ref{ham}.
\subsection{Monte-Carlo simulation}
\label{sec:mc}
In  statistical mechanics, for a system that is in equilibrium with its surrounding environment with temperature $T$, 
the expectation of an observable quantity $\mathcal{O}$ is given by:
\beq
\braket{\mathcal{O}}=\sum_{X} P_{eq}(X)\mathcal{O}_X = \frac{1}{\mathcal{Z}} \sum_{X} \mathcal{O}_X \, e^{-\beta E_X},
\eeq
where $\braket{\cdots}$ indicates the thermal average, $E_{X}$ and $P_{eq}(X)$ are the total energy of the system 
and the probability of being system in the microstate $X$, respectively.
In this expression, $\mathcal{Z}$ is called the partition function ($\mathcal{Z}=\sum_{X}e^{-\beta E_X}$)
and $\beta=1/k_BT$ ($k_B$ is Boltzmann constant).
In a thermodynamic system, the number of particles is an order of Avogadro number. 
Consequently, the number of microstates exponentially becomes large. 
Therefore, the assessment of the partition function is not possible, especially at the thermodynamic limit, $N\to \infty$. 
Part of this obstacle can be overcome by using MC simulations.

The method of the MC simulation, for calculation of the thermal average of a quantity, 
is random choosing of a subset of system microstates ($M$ states). 
Therefore the thermal average approximately becomes as:
\beq
\braket{\mathcal{O}}\simeq\frac{1}{M}\sum_{X=1}^{M} \mathcal{O}_{X},
\eeq
where each $X$ state is randomly selected corresponding to the probability distribution $P_{eq}(X)$. 
These states can be created through the Markov process.
The Markov process obtains a set of successive states $\{X_1, X_2, \cdots\}$ by 
generating each state from the previous one. To use the Markov process, we need to define a transition probability $\mathcal{T}(X_{i}\to X_{j})$.
The transition probability should have the detailed balance condition: 
\beq
P_{eq}(X_j)\mathcal{T}(X_{j}\to X_i)=P_{eq}(X_i)\mathcal{T}(X_{i}\to X_j),
\label{db}
\eeq
which guarantees the generated states follow the probability distribution $P_{eq}(X)$. 
In the Metropolis  algorithm~\cite{Metro}, the transition probability decomposes to the two components:
\beq
\mathcal{T}(X_i\to X_j)=g(X_i\to X_j)\mathcal{A}(X_i\to X_j),
\eeq
where $g(X_{i}\to X_j)$ is the selection probability of 
a change from $X_i$ state to $X_j$ state and 
$\mathcal{A}(X_{i}\to X_j)$ is the acceptance probability.
 Since the selection probability is usually constant, Eq.~\ref{db} can be rewritten as:
\beq
\frac{\mathcal{A}(X_{j}\to X_i)}{\mathcal{A}(X_{i}\to X_j)}=\frac{P_{eq}(X_i)}{P_{eq}(X_j)}
\eeq
So the acceptance probability is defined as follows in the Metropolis algorithm:
\beq
\label{eq-ac}
\mathcal{A}(X_i\to X_j)=min\bigg[\frac{P_{eq}(X_j)}{P_{eq}(X_i)},1\bigg]=min\bigg[e^{-\beta\Delta E},1\bigg],
\eeq
where $\Delta E$ is the energy difference between $X_i$ and $X_j$ states ($\Delta E=E_j-E_i$).
It can be shown that this definition satisfies the detailed balance condition. 
The Metropolis algorithm is also known as the local
update algorithm because only one system site will change in each MC move.
It means a new state $X_j$ is made from the current state $X_i$ of system.
 
In \espin, a random change in the direction of a magnetic moment will bring the system to the new magnetic configuration.
The new magnetic configuration will be accepted according to Eq.~\ref{eq-ac}. 
The energy of the magnetic configurations is calculated using the user-defined spin Hamiltonian. 
If the new configuration is accepted, the system will pass to this configuration
 otherwise the system will remain in its current  magnetic configuration.
To generate a completely distinct configuration from the previous one,
\espin\ sweeps all sites of lattice and applies the local update Metropolis algorithm to each lattice site.
So in \espin, one sweep of lattice is considered as a MC step. 

\begin{figure}[tb!]
\begin{center}
\includegraphics[width=0.3\textwidth,angle= 0.00,clip]{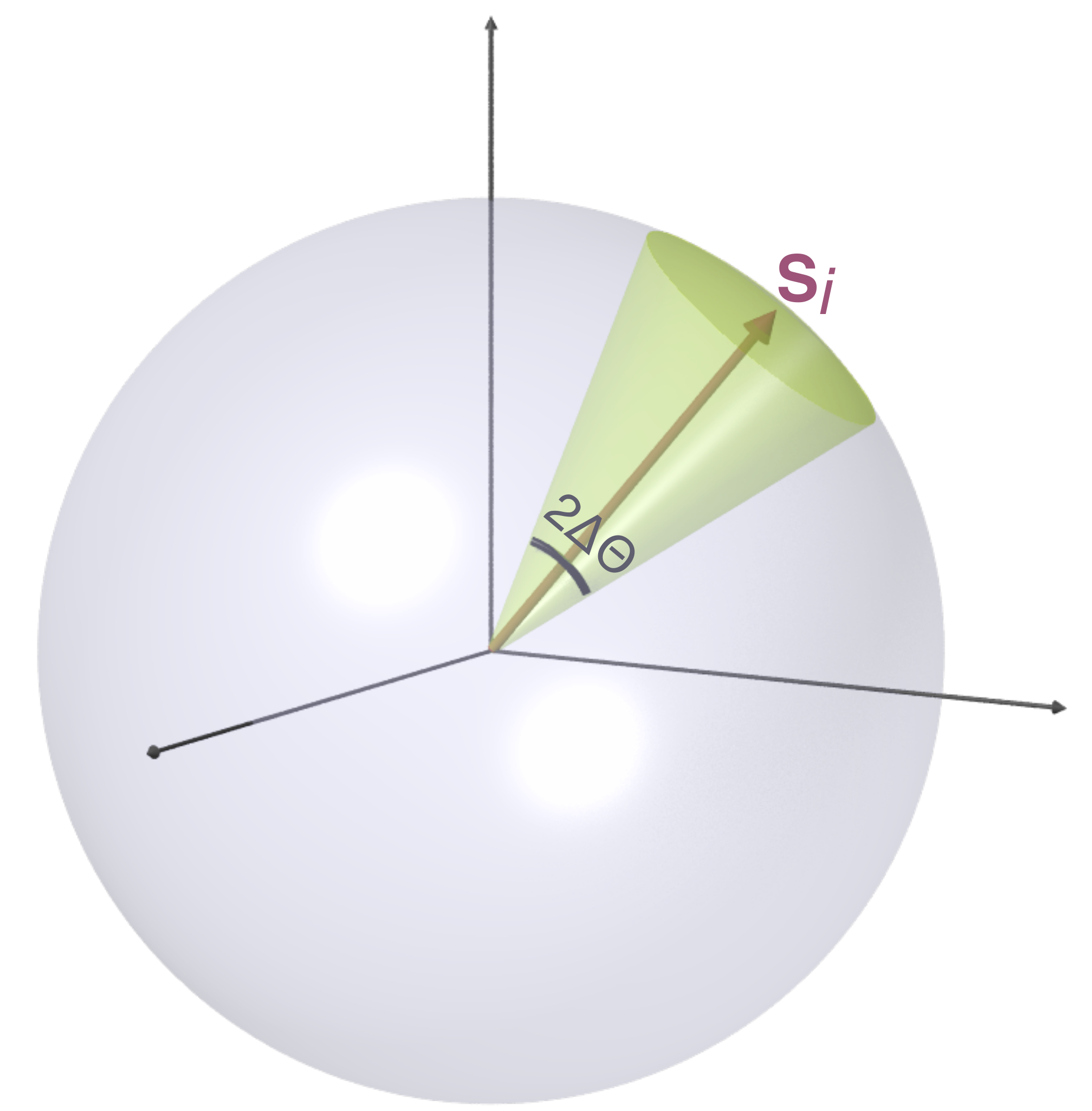}
\end{center}
\caption{(Color online) 
Two algorithms used in MC simulations for local updating of a spin in \espin.
In random mode, the new direction of spin is chosen randomly as a random point in the surface of the sphere, while 
in constrained mode, the new direction of spin restricted to a cone with 
$2\Delta \Theta$ apex angle around the previous direction of spin.
          }
\label{fig:cone}
\end{figure}

For MC simulations at low temperatures and frustrated magnetic systems, 
updating the spin direction by choosing a random spin direction can cause a low acceptance ratio.  
To overcome the problem, \espin\ has an option for constraining the new spin direction.
 In this method, the new direction of spin is chosen randomly inside a cone (Fig.~\ref{fig:cone}). 
The previous direction of the spin becomes the cone axis, and 
the apex angle of the cone ($2\Delta \Theta$)
can be specified in the \espin\ input file. 
The smaller (larger) apex angle leads to a higher (lower) acceptance ratio.
\subsubsection{Parallel tempering algorithm}
The MC local update leads to low acceptance probability 
in complex interacting systems, particularly at low temperatures. 
It is difficult to traverse overall the phase space in such systems due to a rugged energy landscape with a lot 
of local minima separated by high energy barriers.
In this situation, the relaxation time becomes enormously large~\cite{wang-swendsen2005}. Although 
the MC local update still works correctly but 
reaching equilibrium is time-consuming even for a small system. 
Parallel tempering (PT)~\cite{ptmc1,ptmc2,ptmc3}, also known as replica exchange MC, is a technique 
to overcome the slow convergence issue in highly correlated systems.
PT exchanges the configurations of the system at the high temperatures
with configurations at the low temperatures to escape
 energy barriers. The motivation of the method is that, 
 at high temperatures, the whole phase space can be traversed through the local updating. 

In the PT algorithm, the $K$ copies or replicas of the original system 
are simulated 
simultaneously and independently at the different temperatures $\big\{T_1,T_2,\cdots,T_K\big\}$. 
The probability of having the state 
$\{\bm{X}\}=\{(X_1,\beta_1),\cdots,(X_{i},\beta_i),(X_{i+1},\beta_{i+1}),\cdots,(X_K,\beta_K)\}$
in this extended ensemble is given by:
\beq
\label{pt-eq1}
P(\big\{\bm{X}\big\})=\prod_{i}^{K} P_{eq}(X_i,\beta_i)=\prod_{i}^{K}\frac{e^{-\beta_i E_i}}{\mathcal{Z}(\beta_i)},
\eeq
where the ${\mathcal{Z}(\beta_i)}$ is the partition function of the $i$th replica at the temperature $T_i$ ($\beta_i=1/k_BT_i$).
After performing a swapping of $i$th and $i+1$th replicas, the state becomes as\\ 
$\{\bm{X}^\prime\}=\{(X_1,\beta_1),\cdots,(X_{i+1},\beta_i),(X_{i},\beta_{i+1}),\cdots,(X_K,\beta_K)\}$.
Similar to the Metropolis algorithm, 
the PT algorithm should satisfy the condition of detailed balance:
\beq
\label{pt-eq2}
P(\{\bm{X}\})\mathcal{T}(\{\bm{X}\}\to\{ \bm{X}^\prime\})=P(\{\bm{X}^\prime\})\mathcal{T}(\{\bm{X}^\prime\}\to\{ \bm{X}\}).
\eeq
$\mathcal{T}(\{\bm{X}\}\to\{ \bm{X}^\prime\})$ is the transition probability from the state $\{\bm{X}\}$ to state $\{\bm{X}^\prime\}$
 {\it i.e.} accepting probability
of swapping between the $i$th and $i+1$th replicas.
The ratio of transitions can be evaluated by substituting Eq.~\ref{pt-eq1} into Eq.~\ref{pt-eq2}:
\beq
\frac{\mathcal{T}(\{\bm{X}\}\to\{ \bm{X}^\prime\})}{\mathcal{T}(\{\bm{X}^\prime\}\to\{ \bm{X}\})}=e^{\Delta\beta\Delta E},
\eeq
where $\Delta E=E_{i+1}-E_i$ is the energy difference of two replicas and 
$\Delta\beta=1/T_{i+1}-1/{T_i}$ is the difference between their inverse temperatures.
Therefore, the transition probability can be chosen as:
\beq
\mathcal{T}(\{\bm{X}\}\to\{ \bm{X}^\prime\})=min\bigg[e^{\Delta\beta\Delta E},1\bigg]
\label{pt-eq3}
\eeq

Generally, a PT simulation consists of two parts: 
 Metropolis updating of each replica, and swapping the replicas.
After some MC moves (Metropolis updating) on each replica, 
a swapping of replicas at adjacent temperatures is attempted 
according to the PT transition probability (Eq.~\ref{pt-eq3}).
Because the accepting probability decreases exponentially with $\Delta\beta$,
usually, adjacent temperatures are used for swapping. 

In PT simulations, the choice of the temperatures for replicas is critical~\cite{opt_temp2}. 
\espin\ has four options for temperature series:
 a uniform grid on temperatures, a uniform grid on inverse temperatures,  a logarithmic grid on temperatures, and an array of temperatures that are set manually.
The aim of manually setting the temperatures is to optimize PT simulation by, for example, choosing temperatures 
in a way that the swapping ratio between neighboring replicas remains almost constant during the simulation~\cite{opt_temp2}.
\section{\espin\ packagae}
\label{program}
\subsection{Capabilities of \normalfont \espin}
{\em Spin model Hamiltonian:}
 Interactions of Heisenberg exchange, bi-quadratic, \dm, single-ion 
 used to build spin model Hamiltonian. 
The spin-glass model can be described in \espin~by 
random-choosing of  Heisenberg exchanges corresponding to Gaussian probability distribution.
 Periodic and open boundary conditions are permitted on the borders of supercell for the interaction of spins.\\

{\em Initialization of the spin system:}
 To initialize the spin directions at the beginning of MC simulation, 
\espin\ has three options: random spin state, ferromagnetic state, and reading spin directions from a file ({\tt name\_sconfig.dat}). 
The random spin state option can be initialized by either user-defined seed numbers or random seed numbers. \\

{\em MC updating:}
\espin\ has both local update Metropolis and PT algorithm. For the local update of spins, there are two options. 
One is to choose the next spin direction as a new random 3-dimensional unit vector in the real space. 
The alternative option is to select the spin direction in an interval around the previous direction, as was explained in Section~\ref{sec:mc}. \\

{\em Visualization:}
 To visualize the crystal, \espin\ creates a {\tt xsf} structure file at the initialization steps.
Then, the crystal structure can be visualized by VESTA~\cite{Vesta} or XCrySDen~\cite{Xcrysden}, using the {\tt xsf} file.
Also, there is a program in the utility directory to ease visualization of spin directions on the lattice sites by XCrySDen.
\espin\ generates a Python and Gnuplot file to plot the neutron structure factor.\\

\begin{figure}[tb!]
\begin{center}
\includegraphics[width=0.5\textwidth,angle= 0.00,clip]{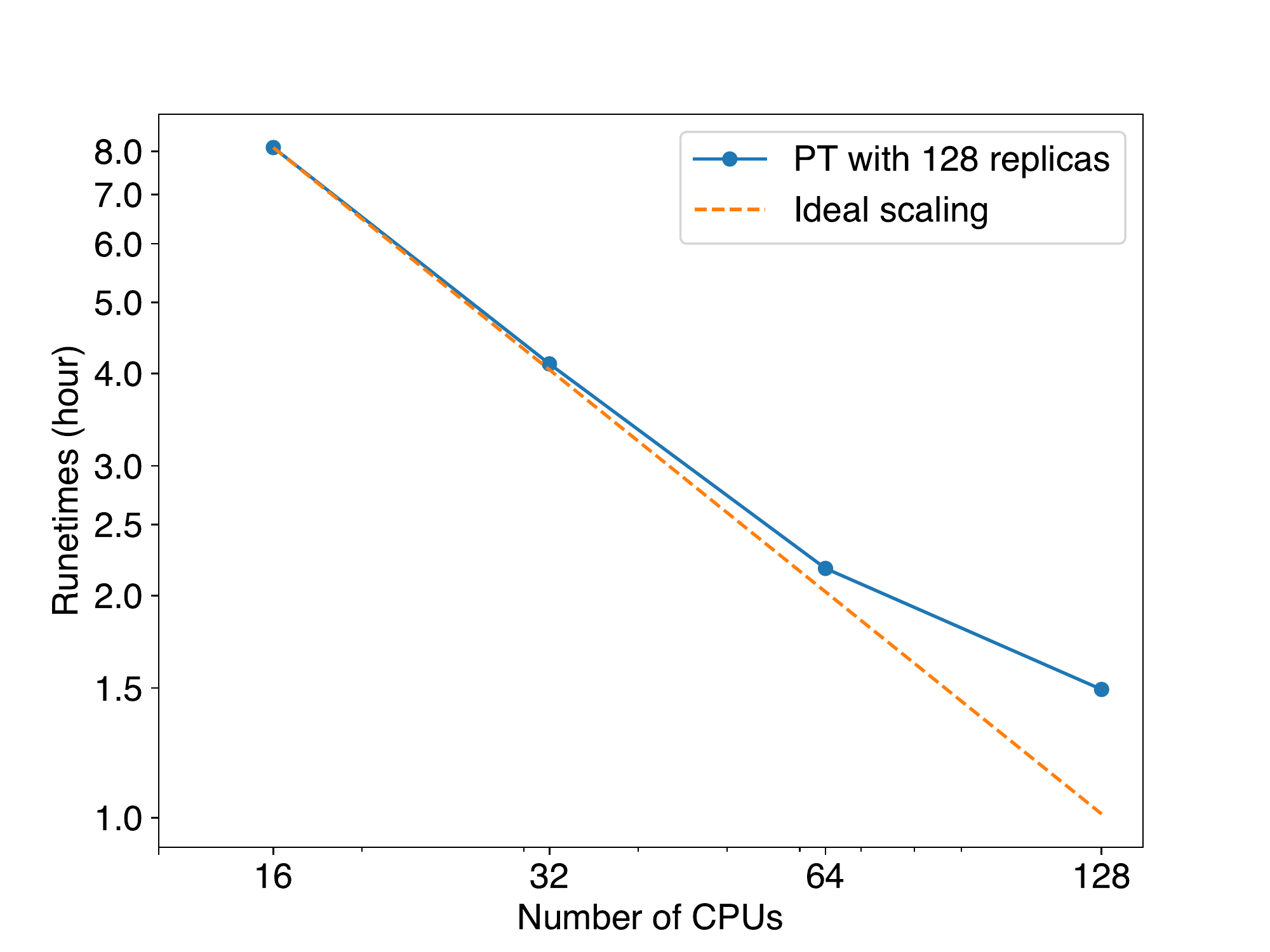}
\end{center}
\caption{(Color online) 
The Scalability of \espin\ with number of CPUs. These calculations have been done using computer nodes with two 8 cores (Intel(R) Xeon(R) CPU E5-2690 $@$ 2.90GHz).
The spin system is a trigonal lattice with two lattice sites, and the size of supercell is set to $15\times15\times15$. The number of temperature points (or replicas) is set to 128.
The total number of MC steps are $10^6$ (so, the total number of local updating is $6.75\times10^{9}$). 
PT swapping happens after every 10 MC steps. 
          }
\label{fig:runtime}
\end{figure}

{\em Parallelization:} The parallelization over temperatures is implemented in \espin.
The number of temperatures should be dividable to the number of processors in the local update algorithm, 
while there is no limitation for the PT algorithm. Fig.~\ref{fig:runtime} illustrates how \espin\ scales with the number of CPUs.
The figure demonstrates that the runtime of PT simulation relatively scales with the number of CPUs. 
\\

{\em Magnetic properties:} \espin\ can calculate the following quantities: 
the total magnetization, magnetic specific heat, susceptibility, fourth-order Binder cumulant for the energy and magnetization, 
staggered magnetization, the probability distribution function of the magnetization, and staggered magnetization, and the neutron structure factor. 
\espin\ can also compute the user-defined order parameter, {\it i.e.},
summation of projections of spins into the user-defined directions.
Consequently, \espin\ calculates the related quantities to the user-defined order parameter such as susceptibility,
 fourth-order Binder cumulant, and probability distribution function of the order parameter.
Additionally, \espin\ can estimate the histogram of the energy either as simple or by the Methfessel-Paxton smearing function.
 In \espin, errors of quantities can be estimated through the binning analysis method~\cite{binning}. 
\subsection{Installation of \normalfont \espin}
\espin\ can be downloaded as a tar.gz compressed file. By the following command, the code will be uncompressed:
\\\\
\hspace*{0.3cm} {\tt \$ tar -xzvf ESpin.tar.gz}
\\\\
This command builds a directory that consists of the source code, example files, etc.
\espin\ is written in Fortran 90, so principally it can be compiled by any Fortran 90 compiler. 
The developers have tested the code with two widespread compilers,  
Gfortran and Intel Fortran compiler (IFORT), on several computer clusters. 
 In addition to a Fortran 90 compiler, \espin\ requires BLAS and LAPACK libraries for installation. 
For the compatibility between Fortran 90 compiler and libraries, the user should modify the {\tt make.sys} file in accordance with her system. 
Two make.sys examples, one for IFORT and others for Gfortran, have been placed in the {\tt config} directory. 
\espin\ can be installed as a serial or parallel version by message passing interface (MPI).
Once the modification of {\tt make.sys} file has been completed,
\espin\ will be installed by typing:
\\\\
\hspace*{0.3cm} {\tt \$ make }  
\\\\                
After compilation, the {\tt mc.x} file is created as the executable program of \espin.
\subsection{Running \normalfont \espin}
A file with {\tt .mcin} suffix is the input file of \espin\ for MC simulation.
To build this input file, \espin\ has two initialization steps. For the first initialization step, \espin\ needs an input file
 with file extension {\tt .inp1.mcin} (for the example {\tt name.inp1.mcin}), which contains the information about the unit cell vectors,
 atomic positions, and shell number for the Heisenberg exchange, bi-quadratic and \dm\ terms of Hamiltonian 
(for a detailed description of the input files, a manual document is distributed with \espin).
The command-line option {\tt -inp1} must be used for the first initialization step:
\\\\
\hspace*{0.3cm} {\tt \$ mc.x -inp1 name}
\\\\
By this command, \espin\ reads the {\tt name.inp1.mcin} file and generates the 
 {\tt name.inp2.mcin} file that contains 
the necessary information to specify the parameters of Hamiltonian.
The user must write the parameters of the corresponding interaction between the atoms in the {\tt name.inp2.mcin} file.
After completing the {\tt name.inp2.mcin} file in the second initialization step,
the following command should be executed:
\\\\
\hspace*{0.3cm}{\tt \$ mc.x -inp2 name}
\\\\
This command-line causes \espin\ to read the {\tt name.inp2.mcin} file and write the {\tt name.mcin} as the main input file for 
MC simulation. This input file contains keywords for the MC simulation such as the total number of MC steps for 
 thermalization and accumulation stages, temperatures, unit cell, atom 
 positions, interactions, etc. Once the file has been completed, \espin\ can be run in a single processor, as follows:
\\\\
\hspace*{0.3cm}{\tt \$ mc.x  name}
\\\\
or in multiprocessor mode:
\\\\
\hspace*{0.3cm}{\tt \$ mpirun -np num\_proc mc.x  name}
\\\\
\espin\ performs the MC simulation and writes some general information about running in the {\tt name.mcout} file.
Some of the computed quantities are written in {\tt name.mcout,} {\tt name\_mc.dat} and {\tt name\_pm.dat} files. 
The details of all output files, as well as input files, can be found in the manual document.
\section{Examples}
We put the input and output files of the following examples in the examples directory of \espin. 
In the following examples, we set the magnetic moments of ions equal to 1 ($\mathbf{S}=1$). 
\label{exam}
\subsection{simple cubic}
\begin{figure}[tb!]
\begin{center}
\includegraphics[width=1\textwidth,angle= 0.00,clip]{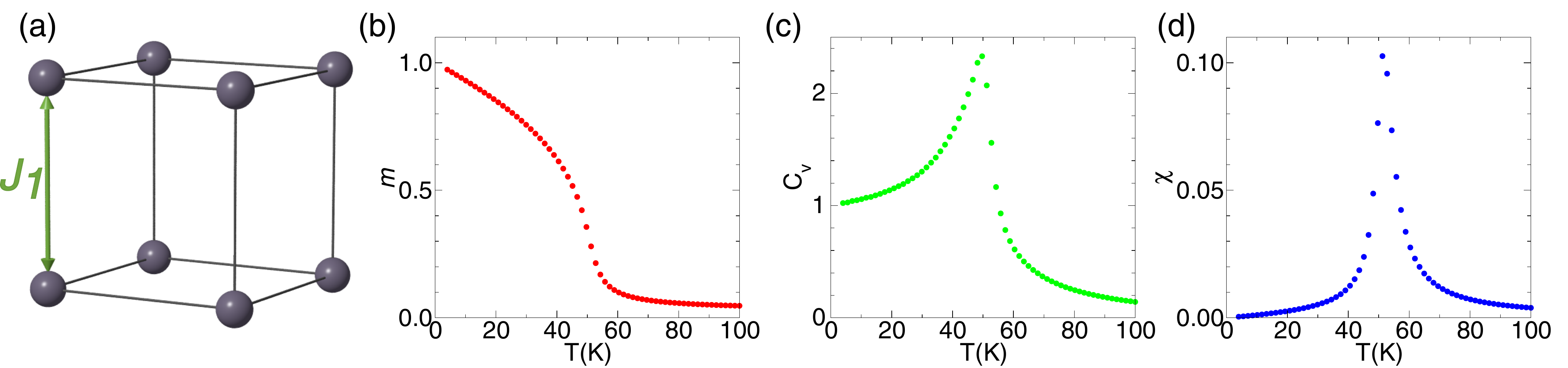}
\end{center}
\caption{(Color online)
(a) cubic lattice. (b) magnetization ($m$) versus temperature. 
(c) magnetic part of specific heat ($\mathrm{C_v}$) versus temperature. (d) susceptibility ($\chi$) versus temperature.
          }
\label{fig:cubic}
\end{figure}
As the first example, we investigate the transition temperature of the simple cubic lattice
with ferromagnetic interactions as much as $J_1 = 3.0$\,meV for the nearest neighbors. 
We set the $1\times10^6$ steps for warming up of the system
and $2\times10^6$ steps for collecting data with
a skip of 5 MC steps between successive data collections to reduce the correlations.
A $10\times10\times10$ supercell consists of 1000 atoms are chosen for MC simulations.
The input files for this simulation are provided in Appendix\ref{app}.
Fig.~\ref{fig:cubic} shows the calculated magnetic thermodynamic properties of this system.
The plots of specific heat and magnetic susceptibility exhibit a peak at $\mathrm{T_c}=$50\,K, 
compatible with the previous study~\cite{cubicmc}, indicating
a transition from ferromagnetic to paramagnetic state with increasing temperature.
\subsection{Antiferromagnetic simple cubic with $J_1=0.26J_2$}
\begin{figure}[h]
\begin{center}
\includegraphics[width=0.95\textwidth,angle= 0.00,clip]{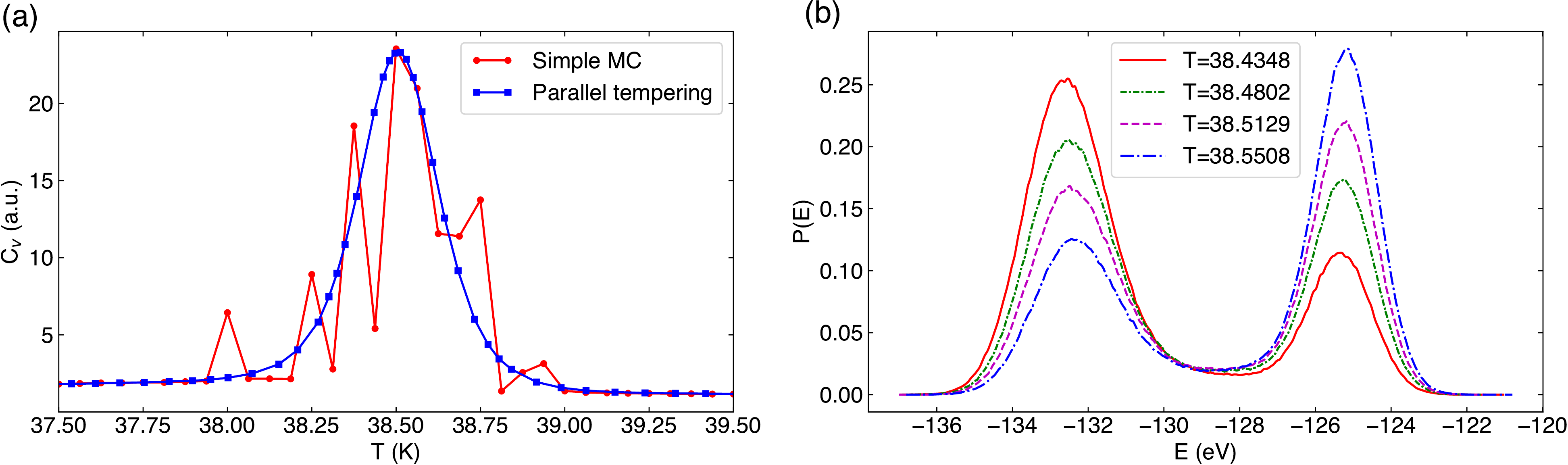}
\end{center}
\caption{(Color online)
(a) specific heat with and without PT simulation (b) the energy histograms near the transition temperature.
          }
\label{fig:cubic_af}
\end{figure}

As the second example, we consider the cubic lattice with antiferromagnetic exchange interactions, $J_1=-10.0$ meV and $J_2=0.26 J_1$. 
The antiferromagnetic next nearest neighbor interaction makes the system frustratedi, and therefore reaching the equilibrium becomes difficult. 
In this example, we show how the PT can help such a system to reach equilibrium faster. 
We set the supercell lattice size $L= 14$, $1\times10^6$ steps for thermalization and data collection with 
a skip of 5 MC steps. 
The specific heat plot with and without PT is presented in Fig.~\ref{fig:cubic_af}(a). 
The smoothness of the heat capacity diagram in the PT method versus its fluctuation 
in the conventional MC method highlights the importance of this method for such systems.
The specific heat peak is around $\mathrm{T_c}=$38.5\,K which is consistent with the result in the literature~\cite{cubic_af}. 
To show the existence of the first-order transition, we also plot the energy histograms, $P(E)$, for several temperatures around $\mathrm{T_c}$ (Fig.\ref{fig:cubic_af}(b)). 
The bimodal distribution of $P(E)$ is the indication of first-order transition~\cite{cubic_af}.
 
\subsection{Pyrochlore $\mathrm{FeF_3}$}
\begin{figure}[tb!]
\begin{center}
\includegraphics[width=1\textwidth,angle= 0.00,clip]{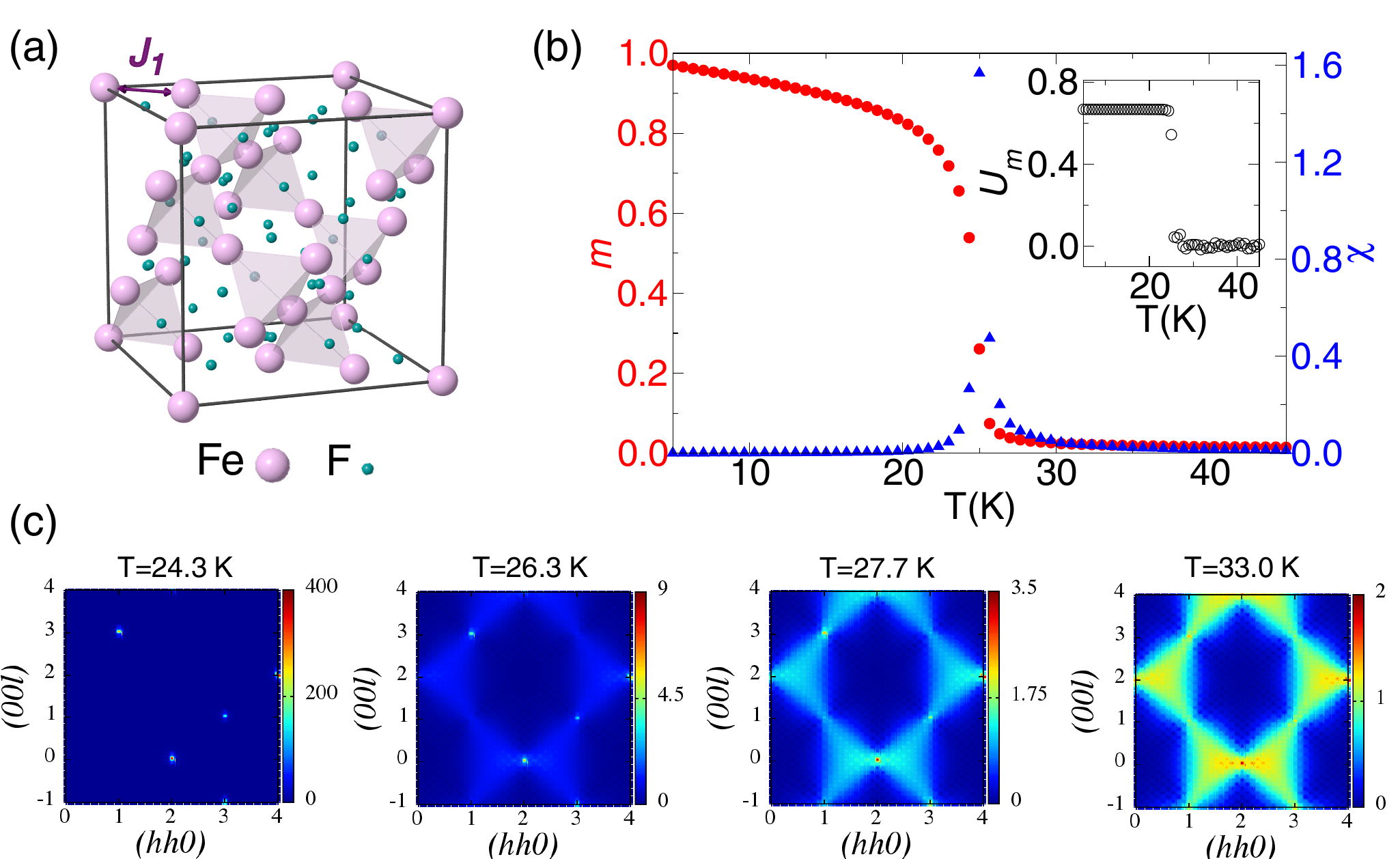}
\end{center}
\caption{(Color online)
(a) crystal structure of $\mathrm{FeF_3}$. (b) AIAO order parameter (circles), $m$, and its respective  susceptibility (triangles), $\chi$, 
versus temperature. Top inside: fourth-order Binder cumulant 
of $m$ versus temperature. (c) Neutron structure factor in $(hhl)$ plane at different temperatures.
          }
\label{fig:fef3}
\end{figure}

$\mathrm{FeF_3}$ has the faced-center cubic (fcc) structure with four Fe$^{3+}$ magnetic ions at each fcc site.
These ions constitute a pyrochlore network of corner-sharing tetrahedra~\cite{Pape1986} (Fig.~\ref{fig:fef3}).
$\mathrm{FeF_3}$ goes to a so-called all-in/all-out (AIAO) magnetic order below $\sim$22\,K~\cite{Ferey1986,Calage1987}. 
We exploit the used spin model Hamiltonian from Ref.~\cite{fef3} that
contains the nearest-neighbors anti-ferromagnetic Heisenberg exchange ($J_1=32.7$\,meV),
 bi-quadratic ($B_1=1.0$\,meV) and \dm\ ($D=0.6$\,meV) interactions.
We use 10$^6$ MC steps for 
both the thermalization and data collection stages at each temperature.
To reduce correlation, we skip every 10 MC steps between successive data collections. 
We choose $10\times10\times10$ for supercell size that contains 4000 Fe atoms. 
The AIAO order parameter is $m=\sum_{i}\mathbf{S}_i.\mathbf{d}_i/N$, where $\mathbf{d}_i$  indicates a 
local unit vector at the $i$th site  which points to the center of the related tetrahedron of the site $i$.

In Fig.~\ref{fig:fef3}, we show computed AIAO order parameter,$m$, 
susceptibility derived from order parameter, 
fourth-order Binder cumulant of the order parameter ($U_m=1-\frac{1}{3}\frac{\braket{m^4}}{\braket{m^2}^2}$), and 
the neutron structure factor at the $(hhl)$ plane. 
The results show a transition to AIAO state at the critical temperature $\mathrm{T_c}\approx24$\,K.
The fourth-order Binder cumulant 
of $m$ is $2/3$ in the AIAO order phase and approaches to zero in paramagnetic order.
The neutron scattering structure factor shows pinch points at the temperatures above the transition temperature. 
Some of these points become the magnetic Bragg points through the transition to the AIAO phase (Fig.~\ref{fig:fef3}). 
Our data is compatible with the results of Ref.~\cite{fef3}. 
It is worth mentioning that the existence of Dzyaloshinskii-Moriya interaction is crucial for AIAO magnetic order~\cite{Elhajal2005}. 
Indeed, in a pyrochlore lattice with only the first nearest neighbor antiferromagnetic exchange,
there is no long-range order~\cite{Reimers1992}.
\subsection{$\mathrm{Bi_3Mn_4O_{12}(NO_3)}$}
\begin{figure}[tb!]
\begin{center}
\includegraphics[width=1\textwidth,angle= 0.00,clip]{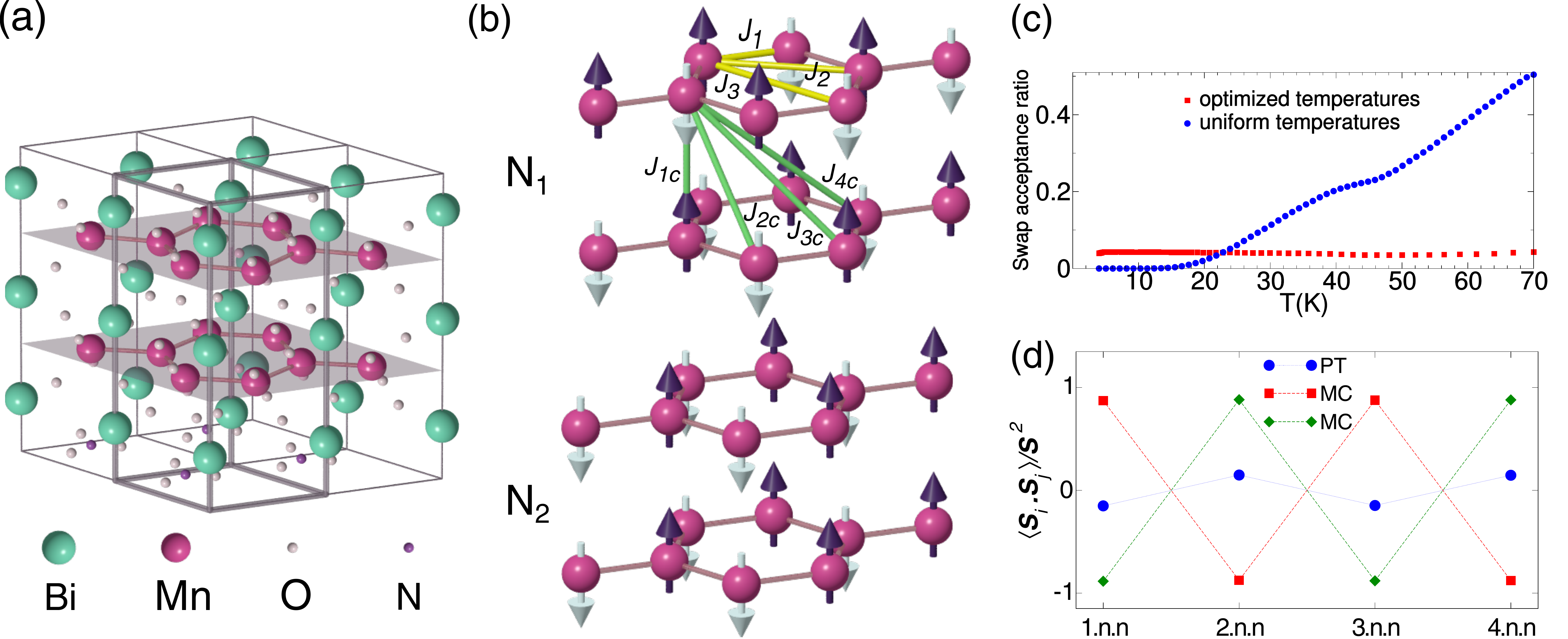}
\end{center}
\caption{(Color online)
(a) Crystal structure of $\mathrm{Bi_3Mn_4O_{12}(NO_3)}$. The primitive cell are shown with thicker
gray lines. (b) N$_1$ and N$_2$ magnetic configurations.
 The Heisenberg exchange coupling constants are shown in N$_1$ magnetic configuration.
 (c) The swap acceptance ratio for uniform and optimized temperatures.  
(d) Inter-layer spin-spin correlation calculated by 
PT and MC calculations. Two different MC results correspond to
 MC simulations with the two different seeds.
          }
\label{fig:bmno}
\end{figure}

In this example, we are focusing on the importance of the PT algorithm in frustrated systems.
$\mathrm{Bi_3Mn_4O_{12}(NO_3)}$ is known as the first frustrated honeycomb lattice,
 wherein Mn$^{+4}$ magnetic ions are arranged in honeycomb bi-layers (Fig.~\ref{fig:bmno}(a)).
It has been shown by MC simulations and calculation of the inter-layer spin-spin correlation\cite{bmno}
 that the lack of magnetic ordering in $\mathrm{Bi_3Mn_4O_{12}(NO_3)}$ is due to the 
being the system at the phase boundary between two N$_1$ and N$_2$ magnetic ordering (Fig.~\ref{fig:bmno}(b)).
Since $\mathrm{Bi_3Mn_4O_{12}(NO_3)}$ goes to the  N$_1$ or N$_2$ magnetic ordering at each run of MC simulation,
in Ref.~\cite{bmno}, the small value of the inter-layer spin-spin correlation
has been gained from averaging over the different runs of MC simulations.
As the PT algorithm is a method for frustrated systems, 
we investigate whether this method can predict the correct value of spin-spin correlation. 
We use the obtained spin model Hamiltonian from Ref. \cite{bmno}, containing the Heisenberg 
terms up to third and fourth neighbors for intra-layer and
inter-layer, respectively ($J_1=10.7, J_2=0.9, J_3=1.2, J_{1c}=3.0, J_{2c}=1.1, J_{3c}=0.5, J_{4c}=0.9$\,meV).
We set $3\times10^6$ steps for thermalization and $7\times10^6$ steps for collecting data with a 10 MC steps skip.
We choose a $30\times30\times1$ supercell containing the 3600 atoms.
For the PT calculations, we specify 64 temperature points between 4-70\,K.

The PT algorithm can be more efficient if the temperature set is optimized~\cite{opt_temp1,opt_temp2,opt_temp3}.
One approach for optimizing temperatures is to choose the temperatures so that the swap acceptance ratio becomes equal for all temperatures.
If the following equation is satisfied for all temperatures:
\begin{equation}
\label{eq:opt}
(\beta_{i-1}-\beta_{i}) [E_{i-1}-E_{i}]=(\beta_{i}-\beta_{i+1}) [E_{i}-E_{i+1}],
\end{equation}
then we expect that the swap acceptance ratio becomes equal for all of the temperatures~\cite{opt_temp3}.
To satisfy Eq.~\ref{eq:opt}, we can run a short MC simulation (for example, with 100000-200000 steps) 
to obtain the system energies for an initial range of temperatures.
Then using interpolation technique and an iterative method, we can estimate optimized temperatures~\cite{opt_temp3}. 
We prepare a program for this purpose in the utility part of the \espin.
Fig.~\ref{fig:bmno}(c) shows how optimized temperatures in the PT algorithm affect the swap acceptance ratio.
Fig.~\ref{fig:bmno}(d) depicts the inter-layer spin-spin correlation obtained from MC and PT 
 simulations at T =4\,K.
As can be seen, the PT algorithm can predict the correct value of spin-spin correlation. 
While in the MC metropolis algorithm, depending on the initial magnetic configuration, the system goes to the one of 
magnetic ordering N$_1$ or N$_2$.
In the uniform temperature set, the swap acceptance becomes zero in the PT algorithm at low temperatures (Fig.~\ref{fig:bmno}(c)). 
As we examine, using a uniform temperature set, and consequently the lack of swapping at low temperatures in the PT algorithm, the system still can trap in one of these magnetic orders.

\section{Conclusion}
\label{con}
In summary, we have presented an open-source computational package called \espin\ 
for obtaining the temperature profile of magnetic properties based on classical MC simulation.
\espin\ was written in Fortran 90 
and can be executed either serial or parallel.
 To make \espin\ user-friendly, the main input file is generated through two initial stages containing fewer keywords.
We described the theoretical basis that \espin\ works on and showed 
how to use \espin.
Finally, we showed the results for some examples, simple cubic, Pyrochlore $\mathrm{FeF_3}$, and $\mathrm{Bi_3Mn_4O_{12}(NO_3)}$
 as a frustrated system. The examples demonstrate some parts of \espin\ abilities in material science research. 
\espin\ is expected to have a major role along with other methods 
such as the first-principles in predicting the magnetic properties of materials.
\section*{Acknowledgments}
We would like to thank Farhad Shahbazi for helpful discussions.
We acknowledge the support of the National
Elites Foundation and Iran National Science Foundation:INSF.
 \appendix
\label{app}
 \section{The sample input file for the first initialization step ({\tt cubic.inp1.mcin})}:
{\scriptsize 
\begin{verbatim}
 Begin  Unit_Cell_Cart
  Bohr
   10.000 0.000 0.000
   0.000 10.000 0.000
   0.000 0.000 10.000
 End Unit_Cell_Cart 

 Begin  Atoms_Frac
  Mn 0.00   0.00   0.00  1.00   
 End  Atoms_Frac
 
 Shells_jij      = 1 
 !! Spin_glass      = .True. 
 !! Ham_bij         = .True. 
 !! Shells_bij      = 1 
 !! Ham_dij         = .Ture. 
 !! Shells_dij      = 1 

 !! Length_unit     = Bohr
 !! Parameter_unit  = Ryd  
\end{verbatim}
}
 \section{The sample input file for the second initialization step ({\tt cubic.inp2.mcin})}:
{\scriptsize 
\begin{verbatim}
  Begin Unit_Cell_Cart
      5.29177211   0.00000000   0.00000000
      0.00000000   5.29177211   0.00000000
      0.00000000   0.00000000   5.29177211
  End Unit_Cell_Cart
  
  Begin Atoms_Frac
   Mn       0.0000000   0.0000000   0.0000000    1.00
  End Atoms_Frac
  
  !! Order_parameter  = .True.
  !! Sfactor          = .True.
  !! Staggered_m      = .True.
  !! Binning_error    = .True.
  !! Spin_correlation = .True.
  !! Energy_write     = .True.

  ## Hamiltonian
  !! Boundary         =  Open 
  !! Ham_Singleion    = .True.
  !! Ham_field        = .True.
  !! Spin_glass       = .True.  !Add the sigma parameters as sig=.. in Parameters_Jij Block

  Begin Parameters_Jij
   t1=  1:t2=  1:sh= 1:Jij= 0.00300!:sig=?????!:d= 5.29177211
  End Parameters_Jij
\end{verbatim}
}
 \section{The sample input file for the MC simulation ({\tt cubic.mcin})}:
{\scriptsize    
\begin{verbatim}
 Begin Unit_Cell_Cart
     5.29177211   0.00000000   0.00000000
     0.00000000   5.29177211   0.00000000
     0.00000000   0.00000000   5.29177211
 End Unit_Cell_Cart
 
 Begin Atoms_Frac
  Mn       0.0000000   0.0000000   0.0000000    1.00
 End Atoms_Frac
 
 tem_start          =   4
 tem_end            =   100
 tems_num           =   64 
 !! tems_mode          = man
 !! tems               = 5.00 10.00 15.00 20.00
 
 !! Pt                 = .True.
 !! Pt_steps_swap      = 10
 
 steps_warmup      =       1000000
 steps_mc          =       2000000
 steps_measure     =            5
 
 initial_sconfig   =       random 
 mcarlo_mode       =       random
 
 supercell_size    =     10    10    10 
 
  ## Hamiltonian
 Begin Jij_parameters
  f1=    0.000000,    0.000000,    0.000000:f2=    1.000000,    0.000000,    0.000000:jij=  0.00300000!:sh=  1!:t1=  1:t2=  1
  f1=    0.000000,    0.000000,    0.000000:f2=    0.000000,    1.000000,    0.000000:jij=  0.00300000!:sh=  1!:t1=  1:t2=  1
  f1=    0.000000,    0.000000,    0.000000:f2=    0.000000,    0.000000,    1.000000:jij=  0.00300000!:sh=  1!:t1=  1:t2=  1
  f1=    0.000000,    0.000000,    0.000000:f2=    0.000000,    0.000000,   -1.000000:jij=  0.00300000!:sh=  1!:t1=  1:t2=  1
  f1=    0.000000,    0.000000,    0.000000:f2=    0.000000,   -1.000000,    0.000000:jij=  0.00300000!:sh=  1!:t1=  1:t2=  1
  f1=    0.000000,    0.000000,    0.000000:f2=   -1.000000,    0.000000,    0.000000:jij=  0.00300000!:sh=  1!:t1=  1:t2=  1
 End Jij_parameters
\end{verbatim}
}

\bibliographystyle{apsrev4-1}
\bibliography{bib}

\end{document}